%% file: main.tex
\documentclass{IEEEtran}
\IEEEoverridecommandlockouts
\usepackage{textcomp}
\usepackage{stfloats}
\usepackage{url}
\usepackage{verbatim}
\usepackage{graphicx}

\usepackage{cite}
\usepackage{amsmath,amssymb,amsfonts}
\usepackage{algorithmic}
\usepackage{graphicx,color}
\usepackage{textcomp}
\usepackage{xcolor}
\usepackage{hyperref}
\hypersetup{hidelinks=true}

\usepackage{subcaption}
\usepackage{stfloats}
\usepackage{url}
\usepackage{verbatim}
\usepackage{subcaption}

\usepackage{tabularx}

\usepackage{algorithm,algorithmic}
\usepackage{cite}
\usepackage{amsmath,amssymb,amsfonts}
\usepackage{algorithmic}
\usepackage{graphicx,color}
\usepackage{textcomp}
\usepackage{xcolor}
\usepackage{hyperref}
\hypersetup{hidelinks=true}

\usepackage{subcaption}
\usepackage{stfloats}
\usepackage{url}
\usepackage{verbatim}

\usepackage{tabularx}

\usepackage{algorithm,algorithmic}
\begin{document}

\title{DESTinE Block: Private Blockchain Based Data Storage Framework for Power System

\thanks{This work was supported by the US Department of Energy 
under award DE-CR0000018.}

}
\author{Khandaker Akramul~Haque\IEEEauthorrefmark{1}, \IEEEmembership{Graduate Student Member, IEEE},
 and Katherine~R.~Davis\IEEEauthorrefmark{1},
\IEEEmembership{Senior Member, IEEE}\\
\IEEEauthorblockA{%
    \IEEEauthorrefmark{1}Department of Electrical and Computer Engineering, Texas A\&M University, College Station, TX, USA \\
  }
}
\maketitle
\thispagestyle{plain}
\pagestyle{plain}

\begin{abstract}
This paper presents DESTinE Block, a blockchain-based data storage framework designed for power systems and optimized for resource-constrained environments, including grid-edge devices such as single-board computers. The proposed architecture leverages the InterPlanetary File System (IPFS) for storing large files while maintaining secure and traceable metadata on a custom blockchain named DESTinE Block. The metadata, comprising the IPFS Content Identifier (CID), uploader identity, administrator verification, and timestamp; is immutably recorded on-chain to ensure authenticity and integrity. DESTinE Block adopts a dual-blockchain abstraction, where the blockchain remains unaware of the IPFS storage layer to enhance security and limit the exposure of sensitive file data. The consensus mechanism is based on Proof of Authority (PoA), where both an administrator and an uploader with distinct cryptographic key pairs are required to create a block collaboratively. Each block contains verified signatures of both parties and is designed to be computationally efficient, enabling deployment on devices like the Raspberry Pi 5. The framework was tested on both an x86-based device and an ARM64-based Raspberry Pi, demonstrating its potential for secure, decentralized logging and measurement storage in smart grid applications. Moreover, DESTinE Block is compared with a similar framework based on Multichain. The results indicate that DESTinE Block provides a promising solution for tamper-evident data retention in distributed power system infrastructure while maintaining minimal hardware requirements.
\end{abstract}


\section{INTRODUCTION}
\input{Sections/introduction}

\section{Algorithm and Analytical Model} 
\input{Sections/model}

\section{Experimental Setup} 
\input{Sections/ex_setup} 

\section{Result} 
\input{Sections/result}

\section{Conclusion} 
\input{Sections/conclusion}

\bibliographystyle{IEEEtran}
\bibliography{ref}

\end{document}

%% file: Sections/introduction.tex
\IEEEPARstart{T}{he} increasing complexity and decentralization of modern power systems have highlighted the need for secure, tamper-evident, and distributed storage mechanisms for critical grid-edge devices such as single-board computers. These devices generate continuous data streams including system logs and measurement values which, if tampered with or lost, can compromise the reliability and accountability of grid operations. Traditional centralized storage approaches face challenges related to single points of failure, limited auditability, and susceptibility to malicious attacks, particularly in remote or resource-constrained environments.

Several blockchain-integrated storage systems have been proposed in recent years. For example, Filecoin \cite{guidi2022evaluating} builds a decentralized marketplace atop InterPlanetary File System (IPFS) \cite{benet2014ipfs} for incentivized file storage using Proof of Replication and Proof of Spacetime, but its computational and economic overhead renders it unsuitable for lightweight or embedded environments. Similarly, Storj \cite{de2021exploring} and Sia \cite{vorick2014sia} utilize blockchain for decentralized cloud storage, though they primarily focus on consumer-level file sharing and cloud alternatives rather than edge-level, domain-specific logging.

Projects like PowerLedger and Grid+ explore blockchain applications in smart grids, but they tend to emphasize energy transactions, market coordination, or demand response rather than decentralized storage for remote terminal unit (RTU) measurements \cite{uddin2023next}. Hyperledger Fabric \cite{androulaki2018hyperledger}, a permissioned blockchain platform, offers high customizability and fine-grained access control, but its overhead and deployment complexity make it impractical for remote edge environments.

To address these limitations, DESTinE Block, a lightweight blockchain-based storage framework specifically designed for secure and verifiable data handling in edge environments of power systems, is proposed. The IPFS is integrated into the system for scalable off-chain file storage, while metadata integrity including file hashes, digital signatures, timestamps, and role identifiers; is maintained on a dedicated DESTinE Block blockchain. This hybrid approach ensures secure and decentralized data management, making it suitable for distributed power system applications. DESTinE Block is derived from the DESTinE (Discrete Event Simulation Tool in Energy) ecosystem, whose core idea is scalability and efficiency for resource-constrained devices \cite{akram_og_destine, haque2024graph}.

Unlike monolithic designs, DESTinE Block separates the data and metadata layers: IPFS stores the full file content, while the DESTinE Block blockchain retains the Content Identifiers (CIDs) and cryptographic proofs. This dual-layer abstraction enhances security and confidentiality, as the actual file content remains inaccessible from the blockchain directly, minimizing data exposure. The system employs a Proof of Authority (PoA) consensus mechanism \cite{manolache2022decision}, requiring co-signatures from two distinct roles, an administrator and an uploader, each possessing independent cryptographic key pairs. This model not only ensures trusted access control but also introduces verifiable accountability, making every data entry traceable to both parties involved in its submission.

DESTinE Block is specifically optimized to operate on low-resource embedded hardware, and its functionality is validated on a Raspberry Pi 5 (ARM64) \cite{rpi5}. This makes it highly suitable for deployment in single-board computers and similar edge devices within smart grids, where computational resources are limited and energy efficiency is critical. A comparable approach was introduced in \cite{cordeiro2025design}, which builds on Multichain \cite{multichain}, and is evaluated against DESTinE Block. While the Multichain-based implementation lacks support for ARM64 architecture, both frameworks were benchmarked on an x86-based device to establish a standard baseline, which was subsequently extended to testing on an ARM64-based Raspberry Pi.

DESTinE Block provides a unique combination of features that address existing gaps in blockchain-based storage solutions for power systems. Its key contributions are outlined as follows:

\begin{itemize}
    \item DESTinE Block is purpose-built for low-power, constrained environments such as single-board computers, and is fully tested on ARM64-based Raspberry Pi 5, ensuring seamless integration at the edge of power infrastructure.
    \item By separating the storage of large files in IPFS from metadata logging on the blockchain, the system achieves stronger confidentiality, modularity, and reduced attack surfaces compared to monolithic blockchain storage designs.
    \item The use of distinct administrator and uploader roles, each with independent key pairs, enforces co-signature-based block creation, ensuring verifiable authorship and protection against single-point compromise.
    \item DESTinE Block supports secure recreation of the blockchain through an encrypted keychain-based persistence mechanism. This approach is resilient to node compromise and functions independently of administrator or uploader credentials, providing additional data integrity safeguards.
    \item Unlike generic blockchain storage platforms, DESTinE Block is explicitly designed for power systems, offering a field-ready, tamper-evident solution for storing operational data and measurements at the grid edge.
    \item DESTinE Block works as a private blockchain framework that deliberately excludes monetary incentives for mining. By adopting a Proof of Authority (PoA) consensus mechanism, the framework ensures transaction validation through pre-authorized entities, thereby eliminating the computational burden associated with Proof of Work (PoW) or Proof of Stake (PoS).
\end{itemize}

The remainder of this paper is organized as follows. Section \ref{model} introduces the algorithm and analytical model of the proposed storage framework. Section \ref{ex_setup} describes the experimental setup employed for performance evaluation. Section \ref{result} presents the results obtained using both an x86-based device and an ARM64-based Raspberry Pi 5. Finally, Section \ref{conclusion} concludes the paper and discusses the future potential and applications of DESTinE Block. The code base of DESTinE Block is available as an open-source repository\footnotemark[1] \cite{khandaker_akramul_2025_17306691}.

\footnotetext[1]{\url{https://github.com/Dr-Kate-Davis-s-Research-Team/destine_block}}

%% file: Sections/model.tex
\label{model}

The backend of DESTinE Block is designed around a GraphQL-based schema, which provides a flexible and efficient application programming interface (API) for interacting with the framework with the help of \emph{Go} programming language \cite{hartig2018semantics, donovan2015go}. GraphQL enables structured queries and streamlined data handling, making it particularly suitable for scalable and resource-constrained environments.

Data uploads within DESTinE Block operate under a Proof of Authority (PoA) consensus mechanism, ensuring that only authorized entities can commit information to the blockchain. Two roles are central to this process: the admin and the uploader. Each role is equipped with a pair of cryptographic keys, one public and one private, that serve as the foundation for authentication and consensus. The uploader initiates the process by preparing data for submission, while the admin must validate and approve the transaction. This dual-key and dual-role mechanism guarantees that no data can be uploaded without mutual agreement, thereby strengthening the integrity and tamper-evidence of the stored records.

In this architecture, the keys serve not only as identifiers but also as integral components of the consensus mechanism, as they are used to digitally sign and verify transactions. Once data is validated, the corresponding Content Identifier (CID) from IPFS is recorded in the blockchain, along with its computed hash and timestamp.

\begin{figure*}[!tbh]
    \centering
    \begin{subfigure}{\textwidth}
         \centering
         \includegraphics[width=\textwidth]{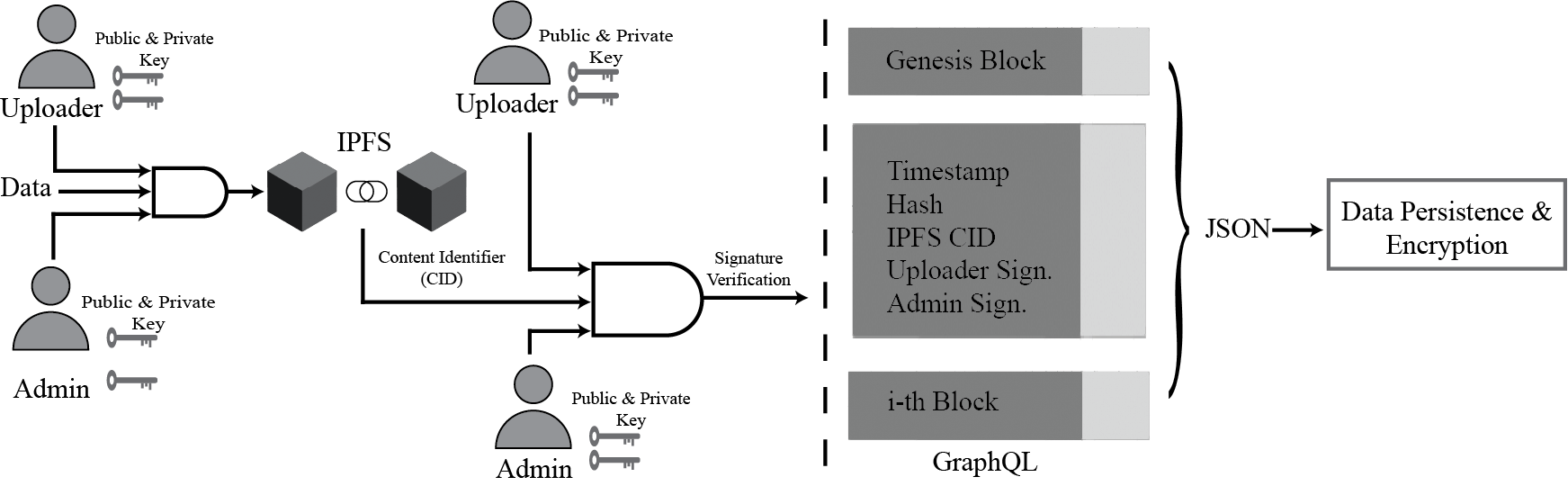}
         \caption{}
     \end{subfigure}
     
     \begin{subfigure}{\textwidth}
         \centering
         \includegraphics[width=\textwidth]{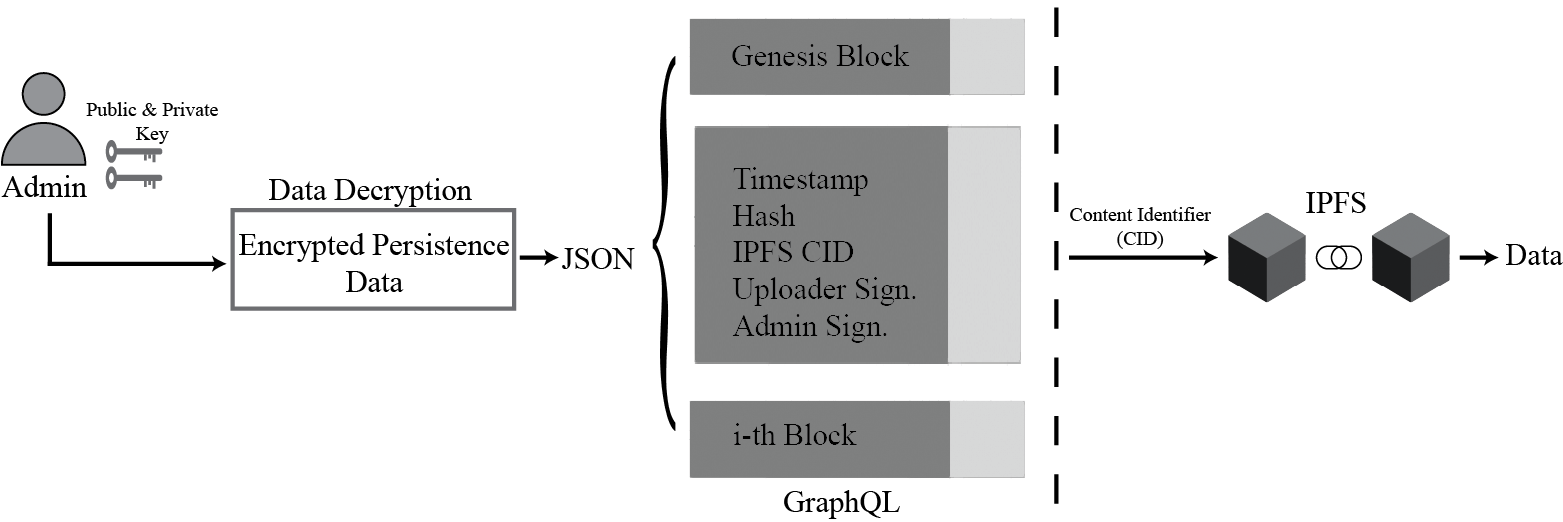}
         \caption{}
     \end{subfigure}
    \caption{Graphical representation of DESTinE Block (a) data upload process; (b) data retreival process.} 
    \label{fig:destine_frame}
\end{figure*}

Fig. \ref{fig:destine_frame}(a) and Fig. \ref{fig:destine_frame}(b) illustrate the workflow of this framework. Figure 1(a) depicts the process of uploading data through the coordinated actions of the uploader and admin. In contrast, Figure 1(b) outlines the retrieval process, where verified information is accessed through the GraphQL API. Together, these processes highlight how DESTinE Block balances security, efficiency, and usability for grid-edge applications.

It is important to note that while data uploads in DESTinE Block result in the creation of new blocks, a subtle distinction exists with respect to the genesis block. The genesis block is generated without the explicit involvement of an uploader, as it does not contain an IPFS CID. In contrast, all subsequent blocks require the participation of both the uploader and the admin. For each new block, the hash of the previous block within DESTinE Block is incorporated into the consensus process, ensuring continuity and immutability. Once validated, a new hash is generated, linking the block securely to the chain and preserving the tamper-evident nature of the framework.

DESTinE Block employs Elliptic Curve Digital Signature Algorithm (ECDSA) keys in conjunction with SHA-256 hashing to ensure the integrity and authenticity of data stored on the blockchain \cite{menezes1993elliptic, gilbert2003security}. In addition, Advanced Encryption Standard in Galois/Counter Mode (AES-GCM) is utilized to generate persistent encrypted data, which can be accessed as a JSON (JavaScript Object Notation) object exclusively within the DESTinE Block framework \cite{dworkinrecommendation}. This design choice enhances portability, allowing secure transfer of data across different edge devices. The proposed data storage framework algorithm, presented in Algo. \ref{algo:des_blo_fra}, outlines this process in detail. A comprehensive analytical explanation of the cryptographic techniques underpinning the algorithm is provided in the following subsection.

\begin{algorithm}[H]
\caption{DESTinE Block Data Storage Framework}
\label{algo:des_blo_fra}
\begin{algorithmic}[1]

\STATE \textbf{Notation:}
\STATE $F$ = file, \quad $\text{IPFS}(F) \to CID_F$ = IPFS content identifier
\STATE $B$ = blockchain storing metadata, $B[i]$ = $i$-th block
\STATE $A=(sk_A, pk_A)$ = Admin keys, \quad $U=(sk_U, pk_U)$ = Uploader keys
\STATE $\mathcal{H}(\cdot)$ = SHA-256, \quad $\mathcal{S}_{sk}(\cdot)$ = ECDSA signing, \quad
$\mathcal{V}_{pk}(\cdot, \sigma)$ = signature verification
\STATE $E_k(\cdot), D_k(\cdot)$ = AES-GCM encryption/decryption with symmetric key $k$

\vspace{0.3cm}

\STATE \textbf{Step 1. Key Initialization}
\STATE Generate elliptic curve keys:
\STATE \hspace{0.5cm} $(sk_A, pk_A) \leftarrow \text{ECDSA.KeyGen}()$
\STATE \hspace{0.5cm} $(sk_U, pk_U) \leftarrow \text{ECDSA.KeyGen}()$

\vspace{0.2cm}

\STATE \textbf{Step 2. Genesis Block Creation}
\STATE $B[0] \gets \{ index=0, prevHash=\emptyset, IPFSHash=\emptyset, owner=A, uploader=U \}$
\STATE $h_0 \gets \mathcal{H}(index \parallel timestamp \parallel prevHash \parallel IPFSHash)$
\STATE $\sigma_U^0 \gets \mathcal{S}_{sk_U}(h_0)$, \quad $\sigma_A^0 \gets \mathcal{S}_{sk_A}(h_0)$
\STATE Store $B[0] \gets (h_0, \sigma_U^0, \sigma_A^0)$

\vspace{0.2cm}

\STATE \textbf{Step 3. File Upload to IPFS}
\STATE $CID_F \gets \text{IPFS}(F)$

\vspace{0.2cm}

\STATE \textbf{Step 4. Block Creation for Metadata}
\STATE $B[n] \gets \{ index=n, prevHash = B[n-1].hash, IPFSHash = CID_F, owner=A, uploader=U \}$
\STATE $h_n \gets \mathcal{H}(index \parallel timestamp \parallel prevHash \parallel CID_F)$
\STATE $\sigma_U^n \gets \mathcal{S}_{sk_U}(h_n)$
\STATE $\sigma_A^n \gets \mathcal{S}_{sk_A}(h_n)$
\STATE Store $B[n] \gets (h_n, \sigma_U^n, \sigma_A^n)$

\vspace{0.2cm}

\STATE \textbf{Step 5. Verification}
\STATE For each $B[i]$:
\STATE \hspace{0.5cm} If $\mathcal{V}_{pk_U}(B[i].hash, \sigma_U^i) \land \mathcal{V}_{pk_A}(B[i].hash, \sigma_A^i) = \text{true}$ then accept block

\vspace{0.2cm}

\STATE \textbf{Step 6. Persistence with Encryption}
\STATE $\text{data} \gets \text{JSON}(B)$
\STATE $C \gets E_k(\text{data})$
\STATE Save $C$ to disk (e.g.: blockchain.json)

\vspace{0.2cm}

\STATE \textbf{Step 7. Retrieval}
\STATE Load $C$ from disk, decrypt $\text{data} \gets D_k(C)$
\STATE Parse JSON $\to B$
\STATE Fetch file: $F' \gets \text{IPFS.Get}(CID_F)$

\end{algorithmic}
\end{algorithm}

\subsection{Cryptographic Primitives}
The cryptographic primitives employed in the DESTinE Block algorithm can be categorized into five core components: ECDSA signing, ECDSA verification, AES-GCM encryption, AES-GCM authentication tagging, and AES-GCM decryption. Together, these elements serve as the fundamental building blocks of the framework.

\subsubsection{ECDSA Signing}

Elliptic Curve Digital Signature Algorithm (ECDSA) is used in DESTinE Block to provide authentication and non-repudiation for each block. The private key of an entity (admin or uploader) is used to generate the signature, while the corresponding public key is later used for verification.

\begin{equation}
d \in \{1, \dots, n-1\}, \qquad Q = dG
\end{equation}

\noindent
Here $d$ represents the private key ($sk_A$ or $sk_U$), while $Q$ is the public key ($pk_A$ or $pk_U$) generated from the elliptic curve base point $G$.

\begin{equation}
e = \text{int}(\mathcal{H}(m)) \bmod n
\end{equation}

\noindent
The message $m$ (e.g., the block hash $h_i$) is hashed using SHA-256, represented by $\mathcal{H}$, and reduced modulo $n$.  

\begin{equation}
k \overset{\$}{\leftarrow} \{1, \dots, n-1\}, \quad R = kG = (x_R, y_R)
\end{equation}

\noindent
A random nonce $k$ (pseudo-random number) is chosen to compute point $R$ on the elliptic curve, which is used to generate the first signature component.

\begin{equation}
r = x_R \bmod n, \qquad 
s = k^{-1}(e + rd) \bmod n
\end{equation}

\noindent
The signature consists of the pair $(r, s)$:
\begin{equation}
\sigma = (r, s) = \mathcal{S}_{sk}(h_i)
\end{equation}

\noindent
This ensures that only the holder of the private key $d$ can generate a valid signature on the block.

\vspace{0.4cm}

\subsubsection{ECDSA Verification}

Verification ensures that a block was indeed signed by the claimed admin or uploader. The verifier recomputes the hash and checks the signature against the public key ($e = \text{int}(\mathcal{H}(m)) \bmod n$).

\noindent
First, the message $m$ is hashed again and reduced modulo $n$.

\begin{equation}
w = s^{-1} \bmod n
\end{equation}

\noindent
The modular inverse $w$ of the signature component $s$ is calculated.

\begin{equation}
u_1 = ew \bmod n, \qquad u_2 = rw \bmod n
\end{equation}

\noindent
Scalars $u_1$ and $u_2$ are then derived.

\begin{equation}
X = u_1 G + u_2 Q = (x_X, y_X)
\end{equation}

\noindent
The verifier reconstructs the elliptic curve point $X$ using the signer’s public key $Q$.

\begin{equation}
\mathcal{V}_{pk}(h_i, \sigma) = \text{true} \;\; \Leftrightarrow \;\; r \equiv x_X \pmod{n}
\end{equation}

\noindent
If the $x$-coordinate of $X$ matches $r$, the signature is valid. This proves that the block was signed by the private key corresponding to public key ($pk$).

\vspace{0.4cm}

\subsubsection{AES-GCM Encryption}

The blockchain state is encrypted before being saved, ensuring that only authorized devices can decrypt it.

\begin{equation}
H = E_k(0^{128})
\end{equation}

\noindent
A hash subkey $H$ is derived by encrypting an all-zero block under the symmetric key $k$.

\begin{equation}
J_0 = IV \,\|\, 0^{31}\,\|\, 1 \quad \text{(for 96-bit IV)}
\end{equation}

\noindent
The initialization vector (IV) is used to generate the pre-counter block $J_0$.

\begin{equation}
\text{ctr}_i = \text{inc}_{32}^{(i)}(J_0)
\end{equation}

\noindent
Counter blocks $\text{ctr}_i$ are derived from $J_0$ for each encryption step.

\begin{equation}
C_i = P_i \oplus E_k(\text{ctr}_i), \qquad i=1,\dots,n
\end{equation}

\noindent
Plaintext blocks $P_i$ (JSON-formatted blockchain data) are XORed with AES-encrypted counters to yield ciphertext $C_i$.

\begin{equation}
C = C_1 \,\|\, C_2 \,\|\, \cdots \,\|\, C_n
\end{equation}

\noindent
The ciphertext $C$ is then stored as a JSON file \texttt{e.g.: blockchain.json}.

\vspace{0.4cm}

\subsubsection{AES-GCM Authentication Tag (GHASH)}

In addition to encryption, AES-GCM provides integrity protection via an authentication tag $T$. The tag is computed using the GHASH function over $GF(2^{128})$, a universal hash function for authenticating AES-GCM encryption.

\begin{equation}
S = \text{GHASH}_H \Big( A, C, \big[\![|A|]\!\big]_{64} \,\|\, \big[\![|C|]\!\big]_{64} \Big)
\end{equation}

\noindent
Here $A$ represents optional additional authenticated data (AAD), and $C$ is the ciphertext.

\begin{equation}
T = E_k(J_0) \oplus S
\end{equation}

\noindent
The tag $T$ is computed by combining the GHASH output with AES encryption of $J_0$.

\begin{equation}
p(z) = z^{128} + z^7 + z^2 + z + 1
\end{equation}

\noindent
Multiplications are carried out in $GF(2^{128})$ modulo the irreducible polynomial $p(z)$.

\begin{equation}
\begin{aligned}
Y_0 &= 0^{128}, \\
Y_i &= (Y_{i-1} \oplus X_i) \cdot H \pmod{p(z)}, \quad i=1,\dots,m
\end{aligned}
\end{equation}

\noindent
This recursion accumulates the authenticated data, ciphertext, and lengths into the final GHASH output.

\vspace{0.4cm}

\subsubsection{AES-GCM Decryption}

Decryption ensures that the ciphertext is not only confidential but also authentic. The authentication tag is recomputed and compared with the received tag.

\begin{equation}
S' = \text{GHASH}_H(A, C, |A|, |C|)
\end{equation}

\begin{equation}
T' = E_k(J_0) \oplus S'
\end{equation}

\begin{equation}
\text{Accept if } T' = T
\end{equation}

\begin{equation}
P_i = C_i \oplus E_k(\text{ctr}_i)
\end{equation}

\noindent
If the recomputed tag $T'$ matches the received tag $T$, the ciphertext $C$ is valid and is decrypted block by block to recover plaintext $P_i$. In DESTinE Block, this yields the blockchain JSON object, which is parsed back into the structure $B$ for use on another edge device.

%% file: Sections/ex_setup.tex
\label{ex_setup}

The experimental evaluation of DESTinE Block was conducted on two distinct hardware platforms: an x86-based laptop and an ARM64-based Raspberry Pi 5. Tab. \ref{tab:dev_spe} summarizes the specifications of the two devices.


\begin{table*}[!t]
\centering
\caption{Device specifications of experimental setup}
\label{tab:dev_spe}
\begin{tabularx}{\textwidth}{|l|X|X|X|}
\hline
\textbf{Device} & \textbf{Processor} & \textbf{RAM/Memory} & \textbf{Storage} \\ \hline
x86-based laptop & Intel i9-12900H & DDR5, 4800 MT/s & NVMe PCIe Gen 4 \\ \hline
Raspberry Pi 5 (ARM64) & Broadcom BCM2712 & LPDDR4X, 4267 MT/s & NVMe PCIe Gen 3 \\ \hline
\end{tabularx}
\end{table*}

The two devices represent substantially different levels of computational capability. The x86-based laptop, equipped with high-performance processing power, faster memory, and next-generation storage, serves as the performance baseline. In contrast, the Raspberry Pi 5 provides a resource-constrained environment, reflecting the type of grid-edge hardware where DESTinE Block is most likely to be deployed.

This deliberate contrast was designed to evaluate two key aspects of the framework: (i) its efficiency and practicality on low-power, resource-constrained devices such as single-board computers, and (ii) whether the use of a more powerful device yields a significant performance improvement in practice. By assessing both extremes, the study provides a comprehensive understanding of DESTinE Block’s scalability and adaptability.

In addition to device-level evaluation, DESTinE Block was compared against a similar blockchain-based storage framework built on Multichain. This comparative analysis offers insights into the relative performance, efficiency, and feasibility of DESTinE Block in contrast to existing technologies.

%% file: Sections/result.tex
\label{result}

\begin{figure*}[!t]
\centerline{\includegraphics[scale=0.68]{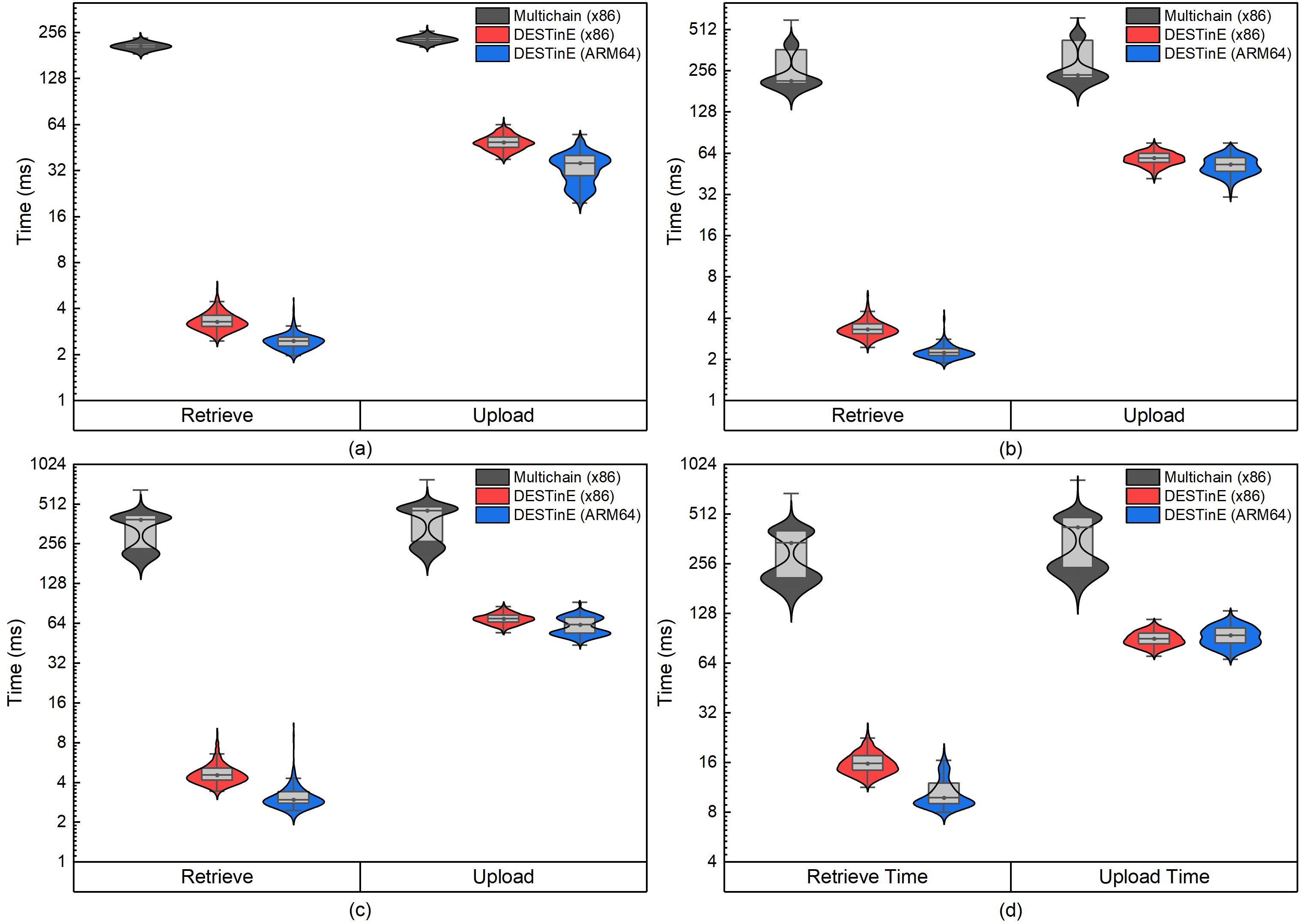}}
\caption{Execution time comparison for file upload and retrieval operations:  Multichain on an x86-based device versus DESTinE Block on x86 and ARM64-based devices. Subfigures show results for (a) 1kB file, (b) 10kB file, (c) 100kB file, and (d) 1MB file.}
\label{fig:des_time}
\end{figure*}

\begin{figure*}[!t]
\centerline{\includegraphics[scale=0.68]{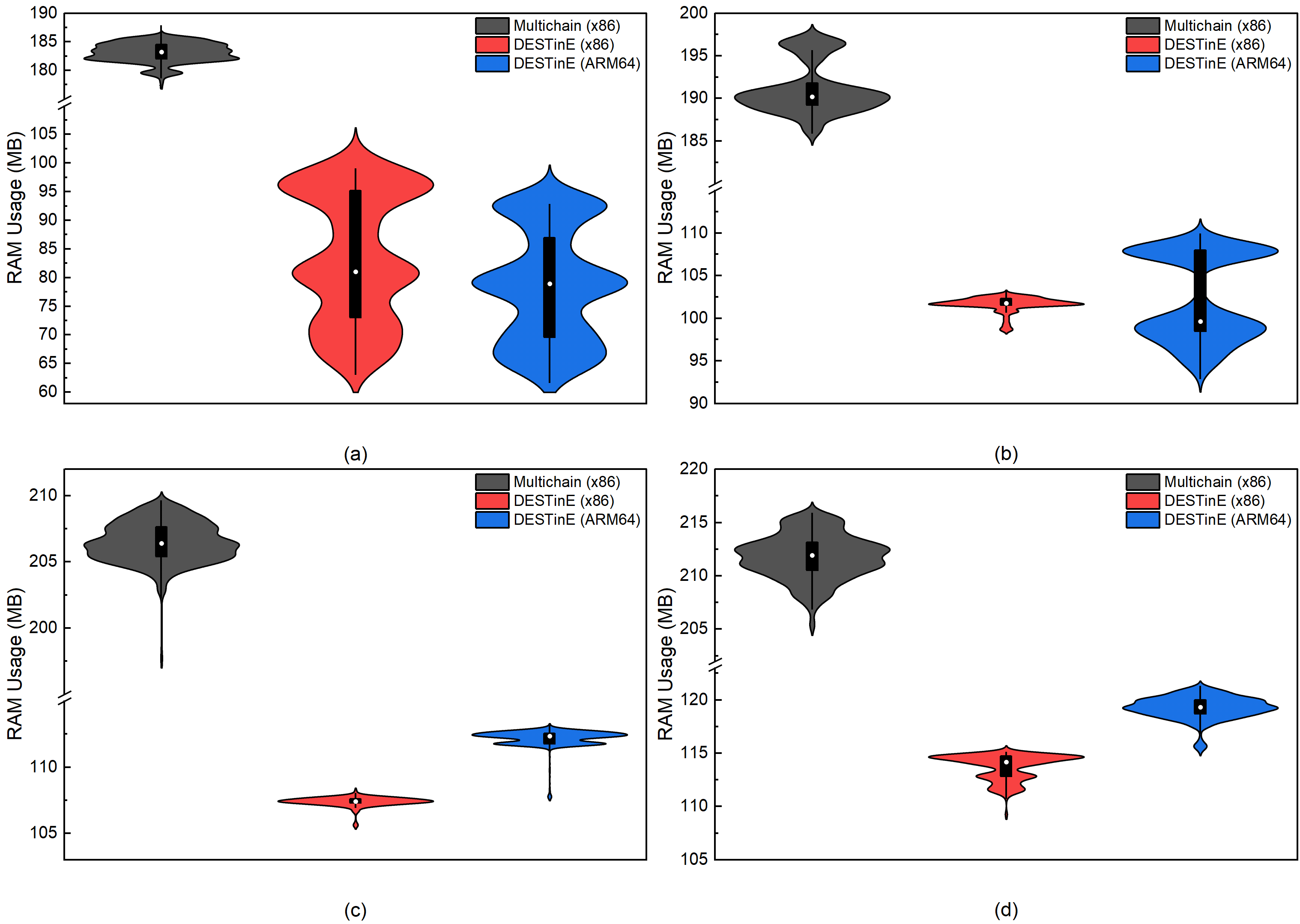}}
\caption{Memory usage comparison for file upload and retrieval operations: Multichain on an x86-based device versus DESTinE Block on x86 and ARM64-based devices. Subfigures show results for (a) 1 kB file, (b) 10 kB file, (c) 100 kB file, and (d) 1 MB file. \emph{Note: memory usage for upload and retrieval is plotted together, as no significant variation was observed between the two processes.}}
\label{fig:des_ram}
\end{figure*}

For the experimental validation of DESTinE Block, binary files of varying sizes, 1 kB, 10 kB, 100 kB, and 1 MB, were selected to emulate different scales of data commonly generated at the grid edge. A binary format was chosen so the contents could be programmatically modified between uploads, enabling controlled variation across trials while the file type remained constant \cite{destine-block-dataset}. Each file was uploaded 1,000 times to the blockchain and subsequently retrieved from both the x86-based device and the ARM64-based Raspberry Pi 5. This procedure was designed to approximate continuous data streams originating from Supervisory Control and Data Acquisition (SCADA) systems \cite{akram_og_destine}.

The primary performance metrics considered were the time required for file uploads and retrievals, as well as system memory consumption during the process. Memory usage was monitored throughout all 1,000 iterations for each file size. To ensure a conservative and consistent assessment, the Virtual Memory Size (VSZ) was used as the benchmark metric for memory consumption, as it provides an upper bound on process memory requirements \cite{gorman2004understanding}.

Moreover, the Multichain-based storage framework is also compared with DESTinE Block. The Multichain-based storage framework was only compared with DESTinE Block on an x86-based device since Multichain does not provide official support for ARM64-based devices, Raspberry Pi 5.

Fig. \ref{fig:des_time} and Fig. \ref{fig:des_ram} represent the time taken for the data uploading-retrieving process and memory usage, respectively, for Multichain only on an x86-based device, while DESTinE Block supports both x86 and ARM64-based devices in the form of violin plot.

Fig. \ref{fig:des_time} demonstrate that DESTinE Block significantly outperforms the Multichain-based framework, with upload times lower by approximately 10 orders of magnitude and retrieval times faster by about 100 orders of magnitude. Across platforms, DESTinE Block on the ARM64-based Raspberry Pi 5 generally achieved faster uploads and retrievals than on the x86 device, though this advantage diminished with increasing file size. In both frameworks, larger files consistently required more time for upload and retrieval. Notably, multimodal patterns in execution time were observed for Multichain-based framework as file size increased, and for DESTinE Block during uploads on the ARM64 platform, likely reflecting the lower hardware specifications of the Raspberry Pi compared to the x86 device.

Fig. \ref{fig:des_ram} indicates that DESTinE Block consistently consumes less memory than the Multichain-based framework, with Multichain requiring nearly twice the memory in most cases. For both frameworks, memory usage increases only slightly as file size grows. DESTinE Block shows comparable memory consumption across the x86 and ARM64-based Raspberry Pi 5 platforms, though a modest upward trend is observed on ARM64 with larger files. Additionally, multimodal patterns in memory usage are evident across all distributions.

Figs. \ref{fig:des_time} and \ref{fig:des_time} revealed multimodality in upload time and RAM usage for DESTinE Block on both x86 and ARM-based devices. To further investigate, we applied a Gaussian Mixture Model (GMM) \cite{steele2010performance}, which represents the probability density of data as a weighted sum of Gaussian components:

\begin{equation}
p(x) = \sum_{k=1}^{K} \pi_k , \mathcal{N}(x \mid \mu_k, \Sigma_k),
\quad \sum_{k=1}^{K} \pi_k = 1,
\end{equation}

\noindent where, $\pi_k$ are mixture weights, $\mu_k$ the means, and $\Sigma_k$ the covariances. Model selection was guided by the Bayesian Information Criterion (BIC):

\begin{equation}
\text{BIC} = -2 \ln(\hat{L}) + k \ln(n),
\end{equation}

\noindent where, $\hat{L}$ is the maximum likelihood, $k$ the number of parameters, and $n$ the sample size. The model with the lowest BIC was selected to balance fit and complexity.

\begin{table*}[!tbh]
\centering
\caption{Summary of Gaussian Mixture Model (GMM) analysis for upload times of DESTinE Block on ARM64 and x86-based devices, evaluated across file sizes of 1 kB, 10 kB, 100 kB, and 1 MB.}
\label{tab:des_time_gauss}
\begin{tabularx}{\textwidth}{|>{\raggedright\arraybackslash}X|c|c|c|>{\centering\arraybackslash}X|>{\centering\arraybackslash}X|}
\hline
\textbf{Variables} & 
\textbf{\begin{tabular}[c]{@{}c@{}}K\\(Components)\end{tabular}} & 
\textbf{\begin{tabular}[c]{@{}c@{}}Log\\Likelihood\end{tabular}} & 
\textbf{BIC} & 
\textbf{\begin{tabular}[c]{@{}c@{}}Cluster\\Mean\end{tabular}} & 
\textbf{\begin{tabular}[c]{@{}c@{}}Cluster\\Variance\end{tabular}} \\ \hline

1 kB DESTinE (ARM64)   & 2 & -3362.02 & 6758.58 & {[}23.48, 36.68{]} & {[}2.17, 39.95{]} \\ \hline
1 kB DESTinE (x86)     & 2 & -3105.35 & 6245.24 & {[}46.88, 53.39{]} & {[}15.21, 27.21{]} \\ \hline
10 kB DESTinE (ARM64)  & 2 & -3492.80 & 7020.16 & {[}47.56, 59.44{]} & {[}26.11, 34.15{]} \\ \hline
10 kB DESTinE (x86)    & 1 & -3283.49 & 6580.81 & {[}59.16{]}        & {[}41.64{]}        \\ \hline
100 kB DESTinE (ARM64) & 2 & -3511.00 & 7077.27 & {[}53.85, 70.58{]} & {[}10.41, 24.23{]} \\ \hline
100 kB DESTinE (x86)   & 2 & -3240.29 & 6515.13 & {[}67.64, 75.038{]}& {[}23.53, 37.14{]} \\ \hline
1 MB DESTinE (ARM64)   & 2 & -3936.48 & 7907.49 & {[}82.21, 100.98{]}& {[}34.63, 62.92{]} \\ \hline
1 MB DESTinE (x86)     & 2 & -3589.00 & 7212.55 & {[}83.76, 96.56{]} & {[}26.59, 51.22{]} \\ \hline
\end{tabularx}
\end{table*}

\begin{table*}[!tbh]
\centering
\caption{Summary of Gaussian Mixture Model (GMM) analysis for RAM/memory usage of DESTinE Block on ARM64 and x86-based devices, evaluated across file sizes of 1 kB, 10 kB, 100 kB, and 1 MB.}
\label{tab:des_ram_gaus}

\begin{tabularx}{\textwidth}{|>{\raggedright\arraybackslash}X|c|c|c|>{\centering\arraybackslash}X|>{\centering\arraybackslash}X|}
\hline
\textbf{Variables} & \textbf{\begin{tabular}[c]{@{}c@{}}K \\ (Components)\end{tabular}} & \textbf{\begin{tabular}[c]{@{}c@{}}Log\\ Likelihood\end{tabular}} & \textbf{BIC} & \textbf{\begin{tabular}[c]{@{}c@{}}Cluster \\ Mean\end{tabular}} & \textbf{\begin{tabular}[c]{@{}c@{}}Cluster \\ Variance\end{tabular}} \\ \hline
1 kB DESTine (ARM64) & 3 & -3369.71 & 6794.68 & {[}67.99, 79.44, 91.36{]} & {[}9.93, 3.63, 5.66{]} \\ \hline
1 kB DESTinE (x86) & 3 & -3467.29 & 6989.86 & {[}69.31, 81.87, 96.46{]} & {[}9.06, 20.96, 2.74{]} \\ \hline
10 kB DESTine (ARM64) & 1 & -3024.58 & 6062.98 & {[}102.11{]} & {[}24.81{]} \\ \hline
10 kB DESTinE (x86) & 2 & -987.84 & 2010.21 & {[}99.26, 101.88{]} & {[}0.36, 0.24{]} \\ \hline
100 kB DESTine   (ARM64) & 2 & -430.29 & 915.86 & {[}111.92, 112.52{]} & {[}0.06, 0.03{]} \\ \hline
100 kB DESTinE   (x86) & 3 & -158.82 & 372.90 & {[}105.20, 107.46, 107.29{]} & {[}2.89, 0.04, 0.16{]} \\ \hline
1 MB DESTine (ARM64) & 2 & -1375.76 & 2786.05 & {[}115.69, 119.38{]} & {[}0.088, 0.74{]} \\ \hline
1 MB DESTinE (x86) & 2 & -1267.67 & 2569.88 & {[}112.60, 114.57{]} & {[}1.28, 0.07{]} \\ \hline
\end{tabularx}

\end{table*}


\begin{figure}[!tb]
\centerline{\includegraphics[scale=0.33]{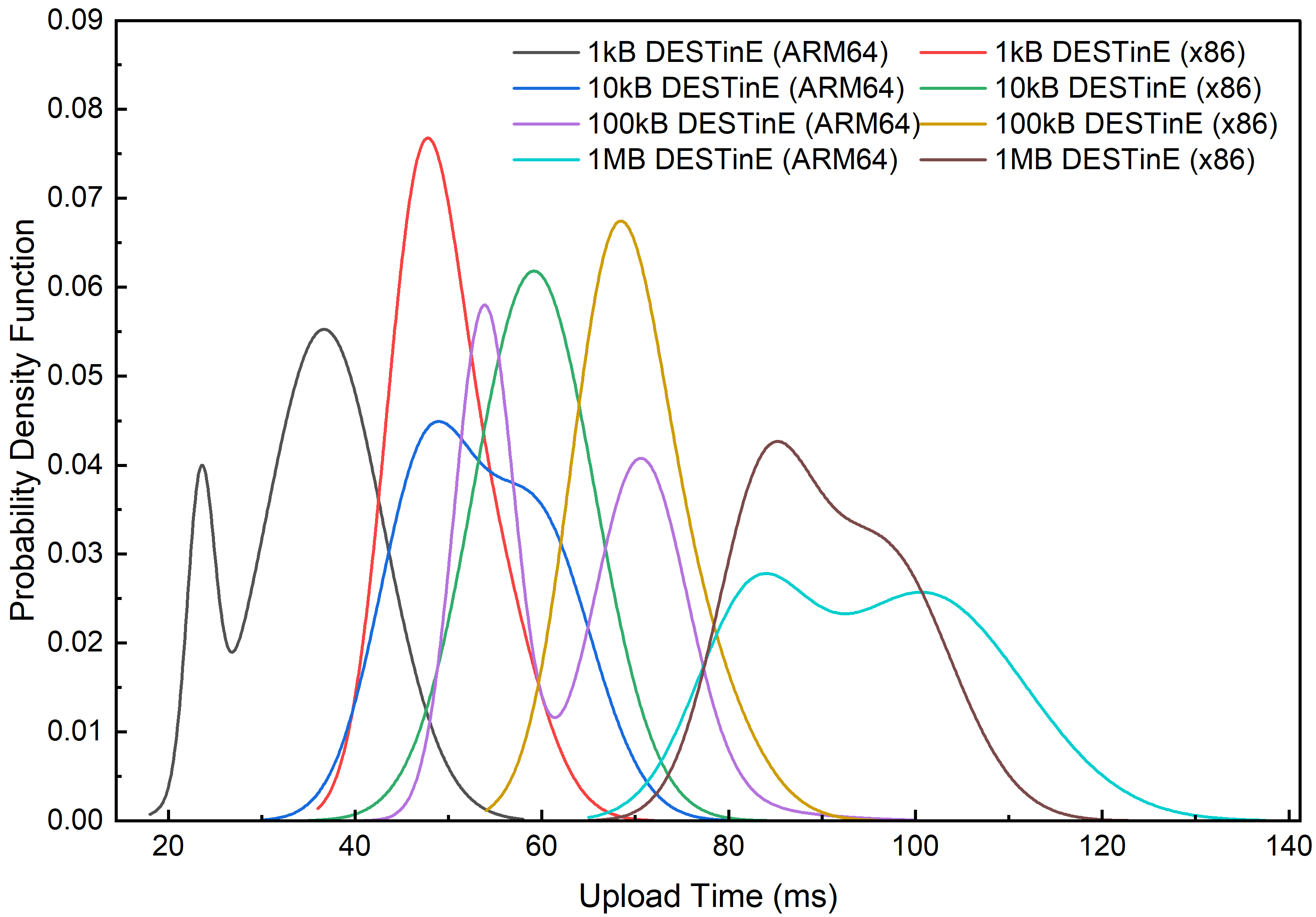}}
\caption{Gaussian Mixture Model (GMM) envelopes of upload time distributions for DESTinE Block on x86 and ARM64-based devices across file sizes (1 kB, 10 kB, 100 kB, and 1 MB).}
\label{fig:des_time_gauss}
\end{figure}

\begin{figure}[!tb]
\centerline{\includegraphics[scale=0.33]{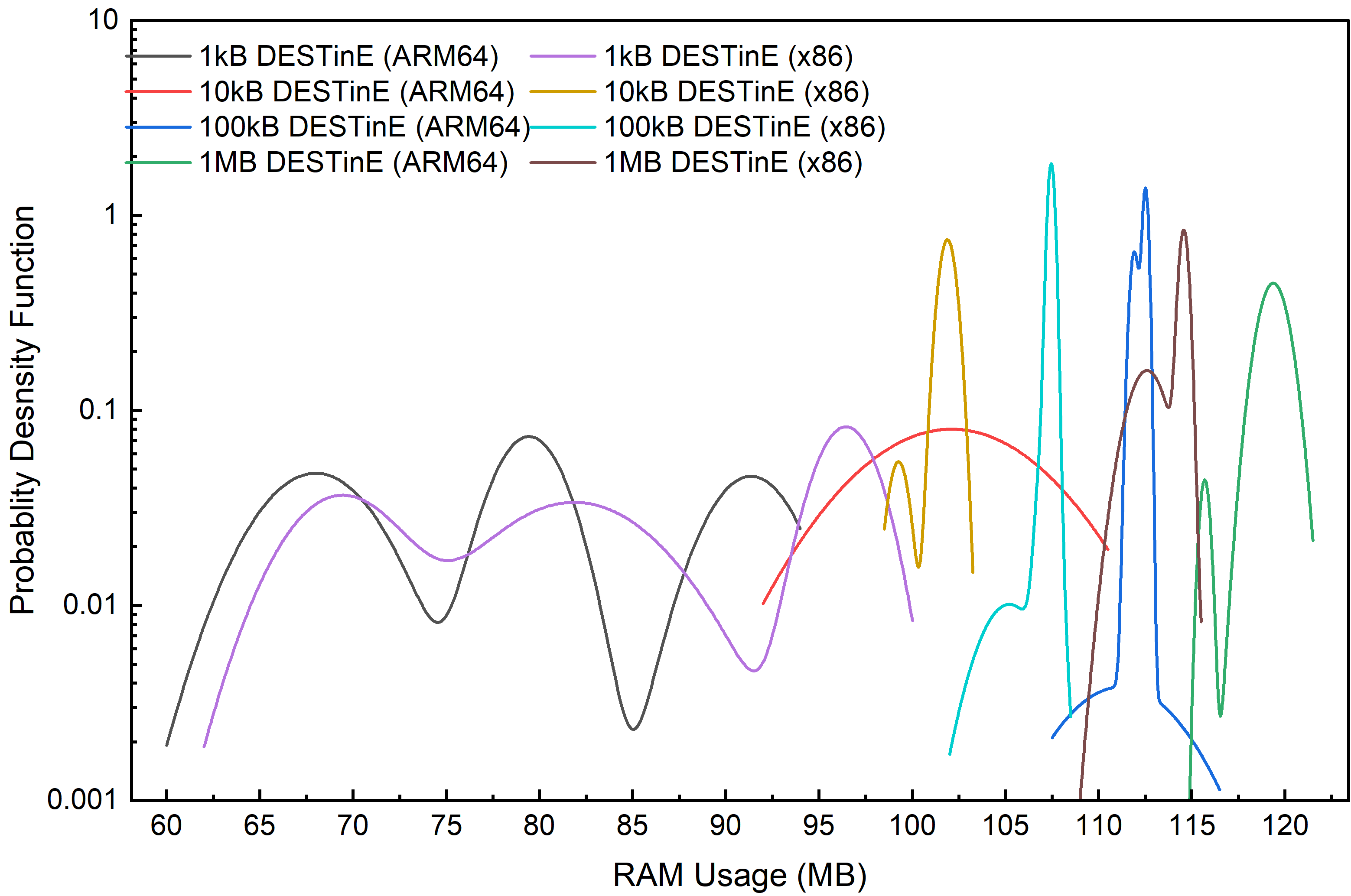}}
\caption{Gaussian Mixture Model (GMM) envelopes of RAM usage distributions for DESTinE Block on x86 and ARM64-based devices across file sizes (1 kB, 10 kB, 100 kB, and 1 MB).}
\label{fig:des_ram_gauss}
\end{figure}

Figs. \ref{fig:des_time_gauss} and \ref{fig:des_ram_gauss} show the GMM envelopes for upload time and RAM usage across file sizes (1 kB, 10 kB, 100 kB, and 1 MB). The results indicate that DESTinE Block on the ARM64-based Raspberry Pi 5 achieves performance comparable to, or in some cases better than, the x86-based device for smaller file sizes (1 kB and 10 kB). While performance on ARM64 declines slightly with larger files, it remains within a similar range to x86. These findings suggest that DESTinE Block can be effectively deployed on resource-constrained devices, such as the Raspberry Pi 5, without significant performance loss. Summaries of the GMM analysis for upload time and RAM usage are provided in Tabs. \ref{tab:des_time_gauss} and \ref{tab:des_ram_gaus}, respectively.


%% file: Sections/conclusion.tex
\label{conclusion}

DESTinE Block presents a secure and efficient blockchain-based data storage framework specifically designed for private, resource-constrained environments. By leveraging Proof of Authority consensus and cryptographic primitives such as ECDSA and AES-GCM, the framework avoids the computational intensity of traditional public blockchains while ensuring tamper-evidence, authenticity, and portability. Experimental validation on both high-performance x86 hardware and the ARM64-based Raspberry Pi 5 confirms that DESTinE Block maintains comparable performance across heterogeneous platforms, with Raspberry Pi 5 often performing equally well for smaller file sizes. Although a slight performance degradation is observed on ARM64 devices with increasing file size, the results remain within acceptable margins, reinforcing the framework’s suitability for grid-edge deployments where efficiency and reliability are paramount. Moreover, the Gaussian Mixture Model analysis of multimodality in upload time and memory usage provides statistical evidence that DESTinE Block remains stable even under varying operational conditions. These findings collectively demonstrate that DESTinE Block can deliver scalable and secure data storage for distributed power system infrastructures without requiring heavy computational resources. For the future, DESTinE Block holds strong potential for integration into proprietary hardware and power system platforms, enabling widespread adoption in critical infrastructure where lightweight, verifiable, and decentralized data management is essential.